# Escape of about five per cent of Lyman-α photons from high-redshift star-forming galaxies

Matthew Hayes[1], Göran Östlin[2], Daniel Schaerer[1,3], J. Miguel Mas-Hesse[4], Claus Leitherer[5], Hakim Atek[6], Daniel Kunth[6], Anne Verhamme[7], Stéphane de Barros[1] & Jens Melinder[2]

[1]Observatoire Astronomique de l'Université de Genève, 51 chemin des Maillettes, CH-1290 Sauverny, Switzerland. [2]Oskar Klein Centre, Department of Astronomy, AlbaNova University Center, Stockholm University, 10691 Stockholm, Sweden. [3]Laboratoire d'Astrophysique de Toulouse-Tarbes, Université de Toulouse, CNRS, 14 Avenue E. Belin, 31400 Toulouse, France. [4]Centro de Astrobiología (CSIC-INTA), PO Box 78, 28691 Villanueva de la Cañada, Madrid, Spain. [5]Space Telescope Science Institute, 3700 San Martin Drive, Baltimore, Maryland 21218, USA. [6]Institut d'Astrophysique de Paris (IAP), 98bis boulevard Arago, 75014 Paris, France. [7]Oxford Astrophysics, Department of Physics, Denys Wilkinson Building, Keble Road, Oxford OX1 3RH, UK.

**The Lyman-α (Lyα) emission line is the primary observational signature of star-forming galaxies at the highest redshifts[1], and has enabled the compilation of large samples of galaxies with which to study cosmic evolution[2–5]. The resonant nature of the line, however, means that Lyα photons scatter in the neutral interstellar medium of their host galaxies, and their sensitivity to absorption by interstellar dust may therefore be enhanced greatly. This implies that the Lyα luminosity may be significantly reduced, or even completely suppressed. Hitherto, no unbiased empirical test of the escaping fraction ($f_{esc}$) of Lyα photons has been performed at high redshifts. Here we report that the average $f_{esc}$ from star-forming galaxies at redshift $z$ = 2.2 is just 5 per cent by performing a blind narrowband survey in Lyα and Hα. This implies that numerous conclusions based on Lyα-selected samples will require upwards revision by an order of magnitude and we provide a benchmark for this revision. We demonstrate that almost 90 per cent of star-forming galaxies emit insufficient Lyα to be detected by standard selection criteria[2–5]. Both samples show an anti-correlation of $f_{esc}$ with dust content, and we show that Lyα- and Hα-selection recovers populations that differ substantially in dust content and $f_{esc}$.**

The hydrogen Lyα emission line, thanks to its high intrinsic luminosity ($L_{Ly\alpha}$)[1], high equivalent width ($W_{Ly\alpha}$)[6,7] and very convenient rest-frame wavelength (1,216 Å), continues to act as a staple tracer of distant star-forming galaxies. However, the Lyα transition is a resonant one, causing photons to scatter in the neutral hydrogen component (H I) of the interstellar medium, so that path lengths to escape may be greatly increased compared to non-resonant radiation. Thus, depending on the content, distribution and kinematics of H I, and the dust content, $f_{esc}$ for a galaxy may fall anywhere within the range of 0 to 1 (refs 8–11). It follows that for the field of Lyα astrophysics to bloom—from using the line to select high-$z$ galaxies, to a physically meaningful diagnostic of star formation—a detailed empirical examination of $f_{esc}$ in cosmological galaxies is essential.

Estimating $f_{esc}$ at high $z$ is extremely challenging, because the requisite supporting data are observationally expensive to obtain. Comparison of star-formation rates (SFR) derived from Lyα with those from ultraviolet continuum can be used to infer $\langle f_{esc} \rangle$ = 30–60% from several studies of $z$ = 2 and 3 Lyα samples[4,5,12]. However, this assumes that the ultraviolet continuum is un-attenuated, and is strongly dependent on models of stellar evolution to provide the respective SFR calibrations[13]. Furthermore, this technique is only valid if star formation has proceeded at a constant rate over the last ~100 Myr. More importantly, restricting analysis to Lyα-selected samples neglects potential star-forming galaxies that do not show Lyα emission. Comparing independently determined Lyα and ultraviolet luminosity functions provides an alternative, but cosmic variance can easily introduce errors of a factor of two (ref. 5), and the aforementioned uncertainties in SFR calibration remain. Using theoretical galaxy formation models, lower values of $f_{esc}$ = 2% (ref. 14) to 10% (ref. 15) have been estimated at $z$ = 3, but these methods suffer from the large number of *ad hoc* parameter assumptions that enter the models. A significant step forward can be taken if Lyα is compared with another, non-resonant hydrogen recombination line (for example, Hα), since both intrinsic strengths are a direct function of the ionizing luminosity. This has been done at $z$ = 0.3, placing $f_{esc}$ = 1–2% (ref. 16) but cosmological application is restricted by the >7 billion years over which galaxies can evolve to $z$ > 2.

With a new, very deep survey using the ESO Very Large Telescope (VLT), we have overcome all of these issues simultaneously. We have performed a blind, unbiased, narrowband imaging survey for Hα and Lyα emission at $z = 2.2$ using custom manufactured filters to guarantee the same cosmic volume is probed in both emission lines (Supplementary Information). Thus, although cosmic variance does affect the number of objects in our survey volume, its effect cancels from any volumetric properties we derive by comparison of the two samples. With observational limits sensitive to un-obscured SFRs of 1.9 solar masses per year ($1.9M_{\odot}$ yr$^{-1}$) in Hα, we identify 55 new Hα emitters[17]. Lyα observations are sensitive to SFR = $0.26M_{\odot}$ yr$^{-1}$ assuming $f_{\rm esc} = 1$ ($f_{\rm esc} = 0.13$ for the faintest Hα emitters), and we identify 38 new galaxies. Lyα and Hα luminosities are shown in Fig. 1 (see also Supplementary Information). Targeting the GOODS-South[18] field, we benefit from some of the deepest broadband optical and infrared data in existence, compiled into public source catalogues[19]. From these we obtain stellar spectral energy distributions (SEDs), which allow us to estimate dust content ($E_{B-V}$) and intrinsic SFR. We find all the Hα galaxies and 21 of the Lyα emitters in public catalogues.

From the Lyα sample we construct the observed Lyα luminosity function, LF(Lyα). Using the same formalism, we derive the first intrinsic Lyα luminosity function, LF(Lyα), from the Hα sample, using measurements of $E_{B-V}$ to correct the Hα luminosity for extinction, and assuming the standard Lyα/Hα line ratio of 8.7 for ionization bounded nebulae[20] ("case B"; see Fig. 2). In this way, we obtain the intrinsic and observed Lyα luminosity densities, the ratio of which leads us directly to a volumetric $f_{\rm esc} = (5.3 \pm 3.8)\%$, with no dependence on cosmic variance, the evolutionary state of the galaxies, or calibration uncertainties. The method is sensitive to the dust correction, as we assume that the same extinction applies to the continuum and to Hα, and thus we perform the same test without correcting Hα luminosities for dust, finding the most conservative upper limit of $(10.7 \pm 2.8)\%$ — a limit free of any model dependency whatsoever. This first key result shows that commonly practised survey estimates of the total Lyα luminosity density at $z \geq 2$ will significantly underestimate its intrinsic value: on average, only 1 in 20 of the intrinsic Lyα photons is accounted for. This may not be surprising at the highest redshifts (that is, above $z = 6$) where an increase in the neutral fraction of the intergalactic medium may cause significant suppression of the Lyα line[21–23], but at $z = 2$–4 this effect is likely to be small, and the photons must be lost in the interstellar medium of individual galaxies. This result is especially consequential, because the integrated luminosity density converts directly to the cosmic rate of star-formation, implying that pure Lyα-based estimates of volumetric SFR are in need of strong upward revision.

To investigate the origin of this underestimate, we examine the individual galaxies. Having modelled the SEDs[24] for all the objects found in the broadband catalogues, we obtain homogeneous derivations of dust extinction ($E_{B-V}$) and SFR for both samples. From the Lyα and Hα luminosities, $E_{B-V}$ estimates and recombination theory, we compute $f_{\rm esc}$ in individual sources (limits on $f_{\rm esc}$ are derived for sources detected in only one line by assigning the $1\sigma$ limiting flux to non-detections). In Fig. 3 we show how $f_{\rm esc}$ correlates with $E_{B-V}$ for our observed galaxies, where we also plot the position of 50,000 synthetic galaxies generated using the 'MCLya' radiation transfer code[25], and the $f_{\rm esc}$–$E_{B-V}$ relationship expected from pure dust attenuation[26]. All the synthetic galaxies fall below the theoretical curve, and every observed galaxy except for one lies within $1\sigma$ of this region. This demonstrates how Lyα photons are preferentially absorbed in the interstellar medium of almost all of our galaxies.

Despite the nonlinearity introduced by the multi-parametric Lyα transfer problem, $f_{\rm esc}$ and $E_{B-V}$ remain clearly anti-correlated. The correlation shows a gradient that is 50% steeper than that predicted by pure dust attenuation: the effective extinction coefficient, $k_{1216}$, is found to be 17.8 instead of 12 for normal attenuation[26]. More striking in Fig. 3 is the lack of overlap and significant offset between the populations: the Lyα and Hα samples are almost disjoint in both quantities. We measure median values of $E_{B-V} = 0.085$ (0.23) and $f_{\rm esc} > 0.32$ (<0.035) for Lyα (Hα) emitters. Furthermore, we find the median SFRs to be very different between the two samples: $3.5 M_\odot$ yr$^{-1}$ for Lyα, and $10.0 M_\odot$ yr$^{-1}$ for Hα. Thus Lyα galaxies are significantly less powerful in forming stars, less dusty, and show higher $f_{\rm esc}$ than Hα galaxies. For Lyα, these estimates are based on the 21 brighter galaxies found in public source catalogues and including the remainder would be likely to increase the disparity between the samples. In $f_{esc}$ the difference is further accentuated by the fact that we are considering lower and upper limits for the Lyα and Hα samples, respectively.

Another significant result is that of 55 Hα and 38 Lyα emitters, we detect only 6 galaxies in both lines. These galaxies straddle the individual distributions, with 5 of the 6 falling within $1\sigma$ of the dust attenuation curve. This is unsurprising when examined in light of the individual samples, but it is unlikely that such a relationship would have been predicted on the basis of the $z \approx 0$ objects of similar luminosity, where little obvious correlation is found between the two line intensities[8,27–29]. The fact that so few Hα emitters are detected in Lyα can be attributed to a combination of two factors. First, the extinction coefficient at Lyα is substantially larger than at Hα ($k_{1216}$ = 12.0 compared to $k_{6563}$ = 3.33)[26]: the median $E_{B-V}$ for the Hα emitters corresponds to a 50% reduction in Hα luminosity but an $f_{esc}$ value of just 7%. Second, the fact that only Lyα scatters serves to exacerbate this, and the grey points in Fig. 3 show how $f_{esc}$ can be reduced to below 1%, even with minuscule dust contents. Indeed, for constant star formation (after the equilibrium time of ~100 Myr) with a 'standard' initial mass function and metallicity, $W_{Ly\alpha}$ is ~80 Å (refs 6, 7), and preferential suppression of Lyα by just a factor of 4 would render a galaxy undetected in the survey. Short-lived burst scenarios increase $W_{Ly\alpha}$ to >200 Å (refs 6, 7), requiring preferential attenuation factors of ~10; these are still easily attainable at very modest $E_{B-V}$ (ref. 11). Similarly, the low number of Lyα sources detected in Hα is explained by the large range of escape fractions exhibited by star-forming galaxies: Lyα selection preferentially finds galaxies with higher $f_{esc}$ values and smaller attenuation in Hα, resulting in line ratios nearer the recombination value and comparatively faint Hα. This pushes the Hα fluxes below our detection limit for most galaxies, despite the very deep Hα data.

Increasing the number of co-incident detections is extremely challenging observationally. Owing to the wide range of relative line intensities, a large range of luminosities needs to be spanned in both lines, requiring each observation to be both wide and deep. This is currently feasible in Lyα, but large (~0.5 degree$^2$) infrared imagers are still non-existent on telescopes of the 8–10-m class. Extending this survey to higher redshift will remain unfeasible until the James Webb Space Telescope comes online.

**Supplementary Information** is linked to the online version of the paper at www.nature.com/nature.

**Acknowledgements** This work is based on observations made with ESO telescopes at the Paranal Observatory under programme ID 081.A-0932. The filter used to capture Lyα was financed by the Erik and Märta Holmberg foundation for astronomy and physics. M.H., D.S. and S.d.B. acknowledge the support of the Swiss National Science Foundation. G.Ö. is a Swedish Royal Academy of Sciences research fellow supported by the Knut and Alice Wallenberg foundation, and also acknowledges support from the Swedish research council (VR). J.M.M.-H. is funded by Spanish MICINN grants CSD2006-00070 (CONSOLIDER GTC) and AYA2007-67965. We thank D. Valls-Gabaud, M. Ouchi, and C. Scarlata for discussions. M.H. thanks the people that made that Christmas at Cerro Paranal memorable.

**Author Contributions** M.H. and G.Ö. conceived the programme and manufactured the custom filter. M.H. observed, and processed and analysed the data. S.d.B. and D.S. wrote tools for analysis of the SED fitting results. A.V. produced the radiation transfer code with D.S. J.M. contributed to the use and processing of auxiliary data. All authors contributed to the interpretation of the data, the research proposal and manuscript preparation.

**Author Information** Reprints and permissions information is available at www.nature.com/reprints. The authors declare no competing financial interests. Correspondence and requests for materials should be addressed to M.H. (matthew.hayes@unige.ch).

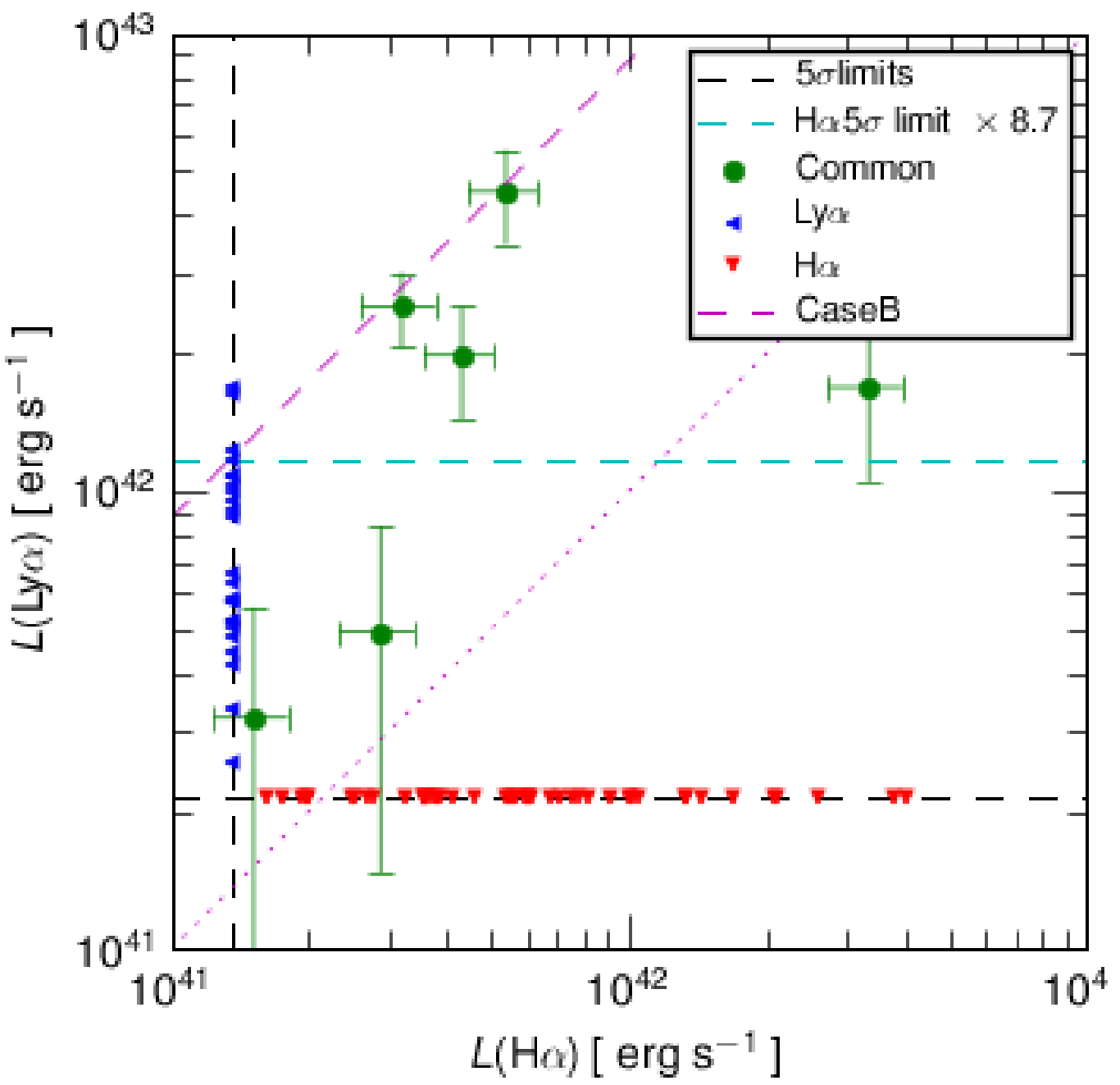

**Figure 1. Observed Hα and Lyα luminosities.** Sources detected only in Hα are shown in red, those detected only in Lyα in blue, and common detections in green. All error bars are 1$\sigma$ photometric uncertainties. Objects undetected in Lyα or Hα are represented as upper limits placed at the detection limits of the Hα or Lyα data, respectively (black dashed lines). The dashed magenta line shows the Lyα/Hα ratio for case B recombination in the absence of dust, and the dotted line shows Lyα = Hα. For a sample

of dust-free galaxies, complete in both lines, all objects should line up on the recombination line, with dust and the effects of radiation transfer serving only to move objects away from the line in the direction of $L_{\text{Ly}\alpha} < 8.7 L_{\text{H}\alpha}$ . The dashed cyan line shows 8.7 times the $5\sigma$ detection limit for Hα applied to the $L_{\text{Ly}\alpha}$ axis. For the case B recombination ratio, all galaxies falling above this line should be detected in Hα (see Supplementary Information for more details). No objects occupy this region of the diagram with significance above $1\sigma$.

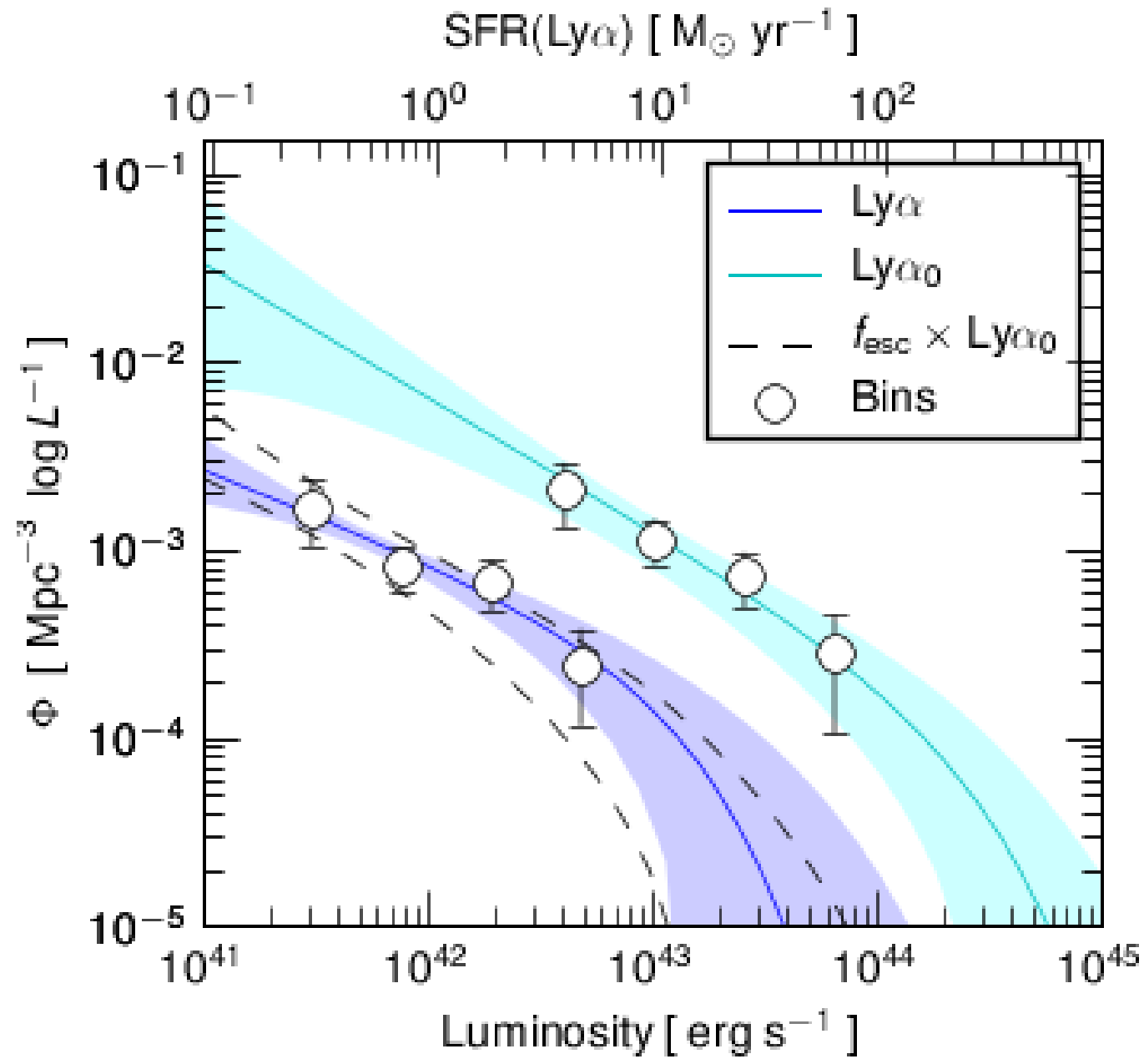

**Figure 2. Lyα luminosity functions.** Φ is the number density of galaxies per decade in luminosity. The SFR labelled on the upper abscissa corresponds directly to the luminosity on the lower. The luminosity function shaded blue, at lower luminosity, shows the observed luminosity distribution, derived from the VLT/FORS1 observations. The function shaded cyan, at higher luminosity, shows the intrinsic luminosity function, denoted LF(Ly$\alpha_0$), derived from the HAWKI observations by correcting the Hα luminosities for dust attenuation, and multiplying by the case B Lyα/Hα ratio of 8.7. Black open circles show the bins of the respective luminosity functions, with vertical error bars representing 68% confidence limits. For the observed LF(Lyα), this error is derived from Poisson statistics and incompleteness simulations alone. For the intrinsic

LF(Lyα), the error bar also includes the error on the dust correction which is randomized on every realization of the Monte Carlo simulation, allowing galaxies to jump between adjacent bins (see Supplementary Information for details). The shaded regions associated with each luminosity function represent the regions of 68% confidence derived from the Monte Carlo. For each realization, both intrinsic and observed LF(Lyα) are regenerated and fitted with the Schechter function; integration over luminosity between 0 and infinity then provides us with the observed and intrinsic Lyα luminosity densities. Volumetric $f_{esc}$ follows directly as the ratio of these two quantities, and is found to be (5.3 ± 3.8)%. Scaling the 68% limits of the intrinsic LF(Lyα) by this fraction in luminosity results in the dashed black lines, which clearly and comfortably encompass the observed distribution.

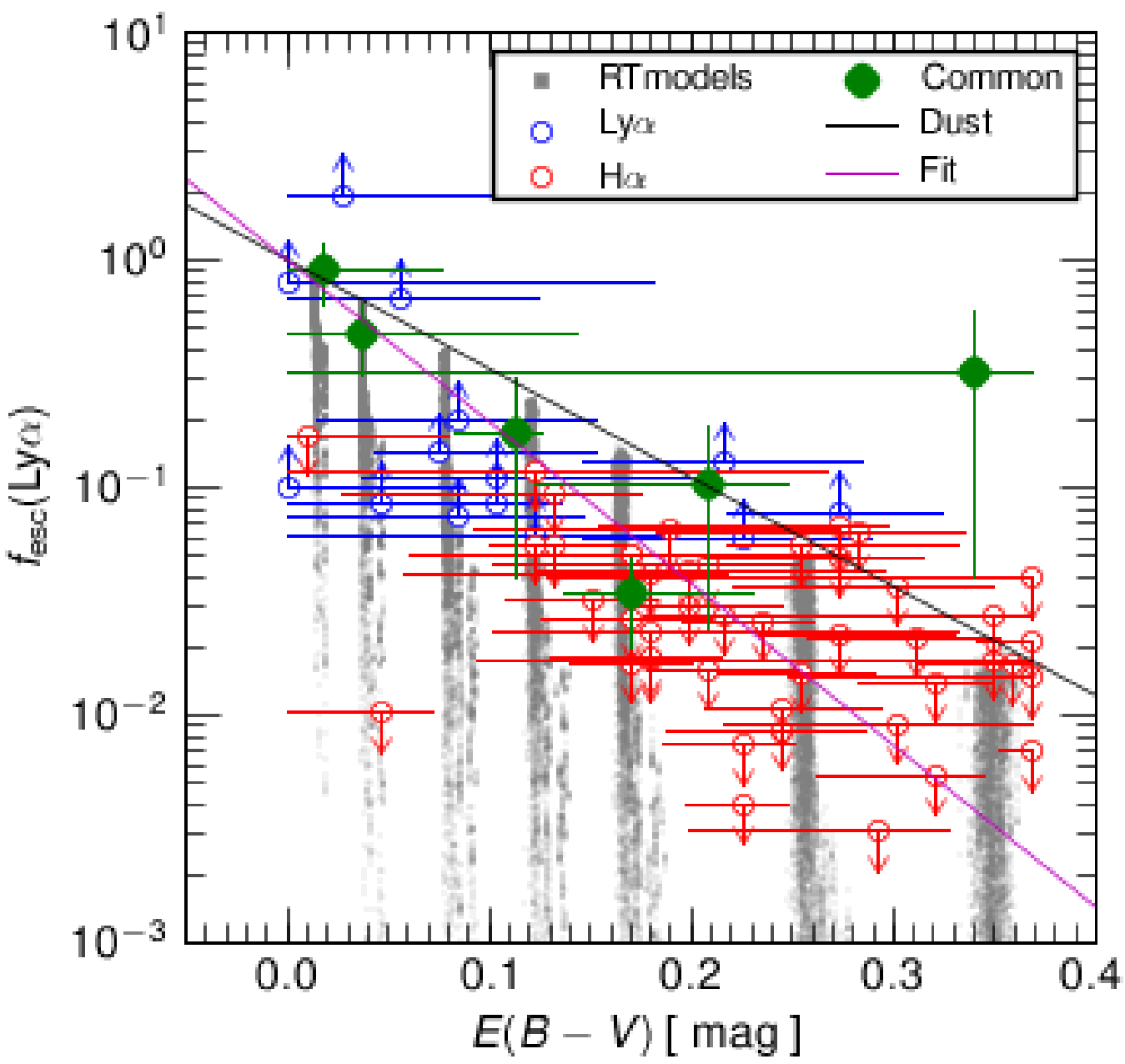

**Figure 3. Escape fraction ($f_{esc}$) and dust attenuation ($E_{B-V}$).** All objects found in the broadband photometry catalogue (that is, for which we can recover $E_{B-V}$) are included. Green shows galaxies detected in both lines, blue shows detections only in Lyα, and red shows detections only in Hα. All error bars are derived from propagation of measurement and model fit uncertainties, and represent 68% confidence. The grey clouds show the positions of 50,000 synthetic galaxies produced using the MCLya radiation transfer code, and are labelled R.T. models. The black line shows the dust

attenuation law of Calzetti, which should be valid in the absence of resonance scattering. The magenta line shows the relation that best fits the observed data points using Schmitt's binned linear regression algorithm[30], a survival analysis algorithm able to account for data points and limits in both directions which does not require a priori knowledge of the distribution of the censored parent population. The number of data points is 67 (55 Hα emitters + 18 Lyα emitters – 6 common detections). Parameterized as $f_{\rm esc} = 10^{-0.4 \times E_{B-V} \times k_{1216}}$, we find the extinction coefficient $k_{1216}$ to be 50% higher ($k_{1216}$ = 17.8 instead of 12.0) than the curve of pure dust attenuation. All but a few data points fall in the region swept out by the radiation transfer code, and demonstrate the significance of the spatial and kinematic structure of the neutral interstellar medium in the transfer of photons. Furthermore it is clear that the areas of the $f_{\rm esc}$–$E_{B-V}$ diagram populated by Lyα- and Hα-selected galaxies are almost disjoint: the Lyα sample are significantly less dusty and exhibit higher escape fractions than the Hα sample. This clearly shows how the populations recovered by the respective selection functions are very different, despite the fact that the physics governing the production of the two emission lines is identical.

# Supplementary Information

## Survey redshift matching and the custom bandpass

Critical for this survey is the matching of the cosmic volumes probed by the Lyα and Hα legs. In right ascension and declination this is easily done thanks to the very similar fields—of—view of the *HAWK-I*[31,32] and *FORS1*[33] cameras, although the redshift dimension is more challenging and required the manufacture of a custom narrowband filter. For Hα we adopted the *HAWK-I/NB2090* bandpass ($\lambda_C$=2.095μm; Δλ=0.019μm) and obtained the overall system throughput curve from the ESO *HAWK-I* Instrument Support Team. We shifted this bandpass in wavelength to the domain required to sample Lyα from the same Δ$z$ as *NB2090* samples Hα: effectively multiplying the wavelength axis of the filter by 1216/6563. We procured a 120mm diameter filter with an extremely similar bandpass for the blue-optimised *FORS1* instrument, the normalised system throughput of which, together with the shifted *NB2090* filter, is shown in Figure SI1. These two filters are clearly an ideal match to sample this slice in redshift. The Lyα filter ($\lambda_C$=3880; $\Delta\lambda$=37.0) is designated *NB388* by ESO and throughout this manuscript.

## Observations and data reduction

The targeted field lies in the GOODS-South field[18], centred at $\alpha=03^h32^m32.^s88$; $\delta=-27^d47^m16^s$. For the 7.'5 x 7.'5 field—of—view of *HAWK-I*, the 191Å full width at half maximum of *NB2090*, and a cosmology of $H_0$=70 km $s^{-1}$ $Mpc^{-1}$, $\Omega_\Lambda$=0.7, $\Omega_M$=0.3, this corresponds to a survey volume of 5440 $Mpc^3$. Selection of the GOODS-S field provides us with a wealth of auxiliary data, including the *Chandra X-ray Observatory, Hubble Space Telescope* (*HST*) observations in *BViz*, *ESO Imaging Survey* data in *UBVRIJHK*$_s$, and *Spitzer Space Telescope* data at 3.6, 4.5, 5.8, 8.0, and 24μm. Furthermore our pointing adopts a position angle of –44° and completely encompasses the Hubble Ultra Deep Field[34] giving us the deepest *BVizJH* data in existence for around one eighth of our survey area.

Details of the *HAWK-I* observations, data reduction, target selection, and luminosity function (LF) have already been presented[17]. Briefly, using *NB2090* and $K_s$ imaging, we identify 152 narrowband-excess candidates with equivalent widths above 63.8Å (i.e. 20Å in the restframe). Using the *GOODS-MUSIC* broadband catalogue[19], we confirm 55 of these to be Hα-emitting galaxies at $z$~2.2 based upon spectroscopic and photometric redshifts, and *BzK* colour criteria. The limiting depth of these narrowband data is AB=24.6 (5σ), which corresponds to a line flux of 6.8 x $10^{-18}$ erg $s^{-1}$ $cm^{-2}$. In turn this corresponds to an unobscured star-formation rate of 1.9$M_\odot$ $yr^{-1}$, assuming a Salpeter[35] IMF and solar metallicity[13]. From analysis of the equivalent width distribution of our $z$=2.2 galaxies we determine that our selection criterion of a minimum equivalent width causes us to underestimate the total Hα luminosity density by between 1 and 16%. We construct LF(Hα) which we find to be well fit by as Schechter function with the parameters of $\log(L^*/\mathrm{erg\ s^{-1}}) = 43.22$, $\log(\varphi^*/\mathrm{Mpc^{-3}}) = -3.96$, and $\alpha = -1.72$.

Using *VLT/FORS1* we obtained very deep narrowband imaging data over an almost identical area using the custom *NB388* filter and also in the continuum using the high throughput *U* and *B* broadbands. Data were obtained in visitor mode during dark time (less than 3 days from new moon and with the moon below the horizon under the duration of observations) over the period of consecutive nights beginning 23 to 27 of December 2008. Table 1 shows the characteristics of the three filters, the total integration times, and number of dithered pointings. Nights beginning 23, 25 and 26 were photometric throughout the duration of the observations. Seeing during the observing run was typically good, never exceeding 1.2".

## Table 1 FORS1 observations

| Filter | $\lambda_C$ [Å] | FWHM [Å] | # Exposures | Exp. time [s] | $m_{lim}$ 5σ [ AB ] |
|---|---|---|---|---|---|
| *NB388* | 3880 | 37 | 36 | 60,900 | 26.4 |
| *U* | 3606 | 513 | 6 | 3600 | 26.4 |
| *B* | 4397 | 1030 | 9 | 3490 | 26.75 |

Table note: In the *U*- and *B*- bands, these exposure times and number of pointings correspond to the observations obtained with *VLT/FORS1* during this observing run. The limiting magnitudes, however, are derived from the *VLT/FORS1* images stacked with the data obtained from the *ESO Imaging Survey*, and thus are significantly deeper than could be obtained in the quoted integration times. The *FORS1* observations were obtained in order to deepen, and better homogenise the dataset.

Data were reduced using standard tasks in NOAO/IRAF: bias subtraction, flat-field correction, and sky subtraction were performed. Images were then registered onto a common astrometric grid and co-added. *U* and *B* images were stacked together with the ESO Imaging Survey *U* and *B* images. The astronomical seeing in the final *NB388* frame was 0.84". The *NB388* frame was calibrated using standard star GD50, observed at various airmasses during photometric time. In the final reduced frames we estimate incompleteness with the ARTDATA package in NOAO/IRAF. We model the point-spread function (PSF) of our final images and insert artificial sources at random positions within the image. We test their recovery using the source extraction and photometry software, described in the following section (using the same configuration). Limiting magnitudes in the various images are also listed in Table 1. The 5σ *NB388* limiting magnitude corresponds to a line flux of $7.8 \times 10^{-18}$ erg $s^{-1}$ $cm^{-2}$. For our faintest Hα fluxes, this represents a Lyα escape fraction of 13%.

**Photometry, selection, and catalogue assembly**

We perform all source detection and photometry with SExtractor[36]. Source detection is done in the *NB388* image, requiring a minimum of 5 contiguous pixels and a threshold signal—to—noise (*S/N*) of 5 above the local background. Photometry is done in the combined *U* and *B*—band images using SExtractor in 'double-image' mode, ensuring that apertures are matched between the respective frames. We adopt the MAG_AUTO method for estimating magnitudes. From the two broadband catalogues we perform a power-law interpolation to estimate an effective magnitude (referred to as *UB*) at the wavelength of the *NB388* filter. We define our selection based upon observed equivalent width, which translates directly into *UB-NB388* colour. To make our selection directly comparable to previous narrowband Lyα surveys[2,3,4,5] we require a restframe $W_{0,\mathrm{Ly}\alpha} > 20$Å. We show the colour—magnitude plot for all the detected sources and the selection of candidates in Figure SI2.

We find 38 galaxies that match this criterion, of which six are also found to be Hα emitters. In Figure 1 of the main article we plot the observed luminosities in Hα and Lyα of all the candidates. Objects undetected in either of the lines are assigned a luminosity based on the 5σ detection limit of the appropriate narrowband observation. The dashed magenta line shows the line ratio expected from case B[20] recombination and none of the objects are found to lie above this line with a significance greater than 1σ. The cyan line shows the 5σ Hα detection limit scaled by the case B line ratio for Lyα/Hα: the luminosity above which all Lyα-selected objects should be detected in Hα. 2 of the 32 candidate Lyα emitters undetected in Hα lie above this line. This can be interpreted as either the presence of a clumpy ISM[37,38] or a non-thermal contribution to the production of Lyα (e.g. winds[39] or cooling radiation[40]) although it is important to stress that both objects are consistent with the limit within 1σ photometric error, and most likely are placed there by scatter.

We cross-correlate all of our candidates with the *GOODS-MUSIC*[19] catalogue in order to obtain broadband magnitudes (or limits in the case of non-detections). 21 of the objects are found in the catalogue. Four of the *GOODS-MUSIC* objects have measured spectroscopic redshifts (spec-*z*) of which two are lower redshift interlopers and two are confirmed Lyα emitters with redshifts consistent with the bandpass. For all the candidates (including the four with spec-*z*'s), we measure their photometric redshifts (phot-*z*) the with the *Hyper-z*[24] code, both in its default state and modified with the inclusion of nebular emission lines[41] (see also the following section). Note that for aperture and methodological consistency, we use as input to *Hyper-z* only the magnitudes published in the *GOODS-MUSIC* catalogue; we never mix *GOODS-MUSIC* photometry with our own. Phot-*z* tests naturally confirm that the two spectroscopically confirmed interlopers are indeed interlopers, and that the two $z$=2.2 Lyα emitters are indeed at $z$=2.2. Of the 17 remaining galaxies, we find phot-*z*'s consistent with $z$=2.2 in all cases apart from one; this object is removed from the sample. For a narrowband filter centred at $\lambda_C$=3880Å there are very few emission lines that could possibly contaminate the Lyα sample: only the [OII]λ3727Å at $z$=0.04 (a comparatively tiny cosmic volume) and the CIVλλ1548,1550Å doublet at $z$=1.5 from active galactic nuclei (AGN). Thus it is not surprising that so few interlopers are found in the sample. In order to identify any objects powered by non-thermal nuclear accretion, we cross-correlate our sample with the 1 Mega-second *Chandra* X-ray catalogue[42,43], but find that no X-ray detections within 2.5" of any of our Lyα candidates. We thus define an interloper rate of 3/21 = 14%. We use this quantity later when selecting Lyα candidates not found in the *GOODS-MUSIC* catalogue to include in the derivation of the Lyα luminosity function.

Finally, we have SEDs for 55 Hα emitters and 18 Lyα emitters. Since six of these galaxies are found in both emission lines we have at total of 55+18-6=67 SEDs.

**SED fitting, dust extinction, and star-formation rates**

For the 67 objects with SEDs we are able to perform detailed fits of the spectral energy distribution using the *Hyper-z* code[24]. We fix the redshift to 2.19, and use synthetic stellar evolutionary models[44] assuming exponentially decreasing rates of star formation, $SFR_t = SFR_0 e^{-t/\tau}$ with $\tau$ in the range 0 (simple stellar population) to $\infty$ (constant star formation) and approximately logarithmically spaced between 5 and 3000 Myr. We assume a Salpeter[35] initial mass function and metallicity $Z$=1/3$Z_{\odot}$. *Hyper-z* employs a standard $\chi^2$ minimiser to cover the full parameter space by brute force, and we fit overall normalisation, stellar age, star-formation history and dust extinction using the Calzetti prescription[26], saving the covariance matrix. We derive confidence limits at 1$\sigma$ from analysis of the $\chi^2$ distribution, searching and interpolating across the full 4-dimensional parameter space and estimating the probability distribution function collapsed onto each dimension. Thus we are able to obtain confidence limits on our best fitting parameters. For this *Letter*, however, the only properties made use of are $E_{B—V}$ and star-formation rate. Note that it is this stellar estimate of $E_{B—V}$ that we use to estimate the reddening undergone by nebular photons (Hα), and apply in our estimate of the intrinsic Lyα luminosity. Studies of star-forming galaxies (8 objects) in the nearby universe[26] have found a nebular $E_{B—V}$ measured from the Balmer decrement to systematically lie a factor of 2 higher than $E_{B—V}$ measured from the stellar continuum. However, application of the Calzetti law for dust attenuation[26] should remove this discrepancy as shown by observations of star-forming galaxies at $z$~2 (114 objects) which show no such offset, instead finding a tight one—to—one correlation between dust-corrected SFR(UV) and SFR(Hα)[45].

We compute some average properties for the respective samples of Lyα and Hα emitters: $E_{B—V}$, $f_{esc}$, and SFR. These quantities are found to be rather different with Lyα emitters less dusty, less star forming, and with higher $f_{esc}$ than Hα galaxies. The Hα sample is complete in detections in the *GOODS-MUSIC* catalogue but Lyα sample is not (21/38) and thus the average values of $E_{B—V}$ and $f_{esc}$ may be biased with respect to the complete sample. However, it is known that for continuum-selected samples[46,47] and also Hα[48] that extinction in general increases with increasing star-formation rate or luminosity. Thus the fainter, undetected Lyα emitters are likely to exhibit lower dust extinctions than those from which the sample-averaged values are calculated, increasing the disparity between the average values computed for the Lyα and Hα samples. By association, and in accordance with the trends presented in Figure 3, this would also likely manifest itself as an increase in $f_{esc}$ for the Lyα-selected sample, increasing the separation between the samples in $f_{esc}$.

**Luminosity functions, Monte Carlo simulations, and escape fractions**

We estimate the intrinsic LF(Lyα) by firstly taking the raw luminosities and $E_{B-V}$ estimates for Hα emitting galaxies and correcting $L_{H\alpha}$ for dust attenuation. For reference our raw Hα luminosity function shows excellent agreement with previous measurements at $z$~2 (ref 49). We then multiply each intrinsic $L_{H\alpha}$ by the case B Lyα/Hα ratio of 8.7 (ref 20) to obtain the intrinsic $L_{Ly\alpha}$ for the individual objects. We bin the objects by luminosity in 4 bins to create the "observed intrinsic" LF(Lyα), shown in cyan in Figure 2. Errors on each bin are derived from Poisson statistics. Using the mean $E_{B-V}$ per bin and the factor of 8.7 again, we convert our intrinsic LF(Lyα) bins back to bins defined by observed magnitudes and, again using the ARTDATA package, simulate incompleteness of the *HAWK-I NB2090* data. Incompleteness is propagated back into the intrinsic LF(Lyα) and its errorbars. To this we then fit the Schechter function using a standard $\chi^2$ minimiser and integrate over luminosity between 0 and infinity to obtain the intrinsic Lyα luminosity density, $\rho_{int}(L_{Ly\alpha})$, for our $z$=2.2 volume.

We assemble the observed LF(Lyα) in a similar manner by first including all the 18 confirmed $z$=2.19 Lyα emitters with redshifts consistent with 2.19. From the 17 candidates with continuum too faint to be found in the *GOODS-MUSIC* catalogue, we randomly select galaxies with an 86% chance of inclusion (i.e. based upon the 14% interloper rate). All selected objects are binned in observed $L_{Ly\alpha}$, and incompleteness is simulated in the *FORS1 NB388* frame using ARTDATA as described previously. We again fit a Schechter function and integrate over luminosity to obtain the observed Lyα luminosity density, $\rho_{obs}(L_{Ly\alpha})$. For reference our observed LF(Lyα) is in reasonable agreement with the $z$=3.1 LF[4,5] although does find $L^*$ around one magnitude brighter at the 1σ confidence level. The observed LF(Lyα) is shown in blue in Figure 2. Note again that while our survey volume is small and cosmic variance may be significant, the volume is perfectly matched between the Hα and Lyα legs of the survey and cosmic variance divides away from results based on the the differential comparison of the two LFs.

We present the Schechter parameters for both the observed and intrinsic LF(Lyα) in Table 2. With measurements of the observed and intrinsic Lyα luminosity densities, we define the volumetric Lyα escape fraction as $\rho_{obs}(L_{Ly\alpha})$ / $\rho_{int}(L_{Ly\alpha})$. From the errors on $L_{H\alpha}$ and $E_{B-V}$ for the Hα sample, and $L_{Ly\alpha}$ only for the Lyα sample we randomly regenerate the catalogues and Lyα luminosity functions, and re-perform the above procedure over a 1,000 realisation Monte Carlo simulation. Objects among the 17 candidates not found in the *GOODS-MUSIC* catalogue are randomly re-drawn on every realisation based upon the $z$=2.2 confirmation rate of the objects that were. This simulation yields a volumetric escape fraction of (5.3±3.8) %. Over each realisation we also perform the same calculation without the inclusion of the reddening correction on Hα as a security check to find the absolute maximum volumetric escape fraction. Here we simply calculate $\rho_{obs}(L_{Ly\alpha})$ / [ 8.7 × $\rho_{int}(L_{H\alpha})$ ]. This places the most stringent limit on the volumetric $f_{esc}$(Lyα) of (10.7±2.8)% and, while higher than our primary method by a factor of 2, this derivation contains not one single model dependency.

Since the luminosity densities are obtained by integration from zero to infinity, this represents significant extrapolation in luminosity from the limited range spanned by our survey. To assess the impact of this, we also sum directly the luminosities of our Lyα and Hα emitters and find the fraction of integrated luminosity missed by the survey. We see 68% of the luminosity in Lyα and 65% for Hα and thus, while the range covered by our survey is limited, around 2/3 of the total luminosity is accounted for purely by addition of luminosities; simply summing the fluxes of Lyα and Hα emitters would give very similar results.

The selection function for Lyα emitters rejects all objects with $W_{Ly\alpha,0} < 20$Å and, if those objects contribute a large fraction to the total Lyα luminosity, this will go undetected by Lyα surveys that cut at the canonical value. To investigate this we examine the equivalent width distribution of the Lyα galaxies in our sample, by first fitting an exponential function to the distribution as previously done at redshifts 2 (refs 12, 50) and 3 (ref 4), and secondly by investigating the luminosity distribution cumulative in $W_{\mathrm{Ly}\alpha}$. The exponential form of the $W_{\mathrm{Ly}\alpha}$ distribution has an *e*-folding scale of 76Å, the same as that observed at z=3.1 (ref 4) and slightly higher than the value of 40-50Å observed at z=2, although poor statistics are likely responsible for the discrepancy. Thus between the limits of $W_{\mathrm{Ly}\alpha}$ being independent of Lyα luminosity and independent of continuum luminosity density, we miss between 3 and 22% of total luminosity. Adopting the narrower z=2 $W_{\mathrm{Ly}\alpha}$ distributions this increases to around 30%. However, examining the cumulative luminosity distribution as a function of $W_{\mathrm{Ly}\alpha}$ we find that the underestimate is likely about 20%, which would cause the volumetric escape fraction to increase to around 6%. On the other hand, a very similar result is also found the Hα emitters[12]: around 20% of the total luminosity density is missed, and therefore overall it is likely that this effect largely cancels in the computation of $f_{\mathrm{esc}}$.

**Table 2 Schechter parameters of the Lyα luminosity function**

| | **log $L^*$ [ erg s$^{-1}$ ]** | **log $\varphi^*$ [ Mpc$^{-3}$ ]** | **α** |
|---|---|---|---|
| Observed | 43.16 +/- 0.32 | —3.63 +/- 0.52 | —1.49 +/- 0.27 |
| Intrinsic | 44.47 +/- 0.35 | —3.96 +/- 0.68 | —1.65 +/- 0.33 |

**The radiation transfer models and theoretical Lyα escape fractions**

*MCLya*[11,25] represents the current state—of—the—art software for three dimensional radiation transfer of Lyα and continuum photons. Physics is implemented to include dust scattering and absorption, HI scattering, frequency and angular redistribution, all in arbitrary geometries and velocity fields. All our models are carried out assuming a spherically symmetric, homogenous and co-spatial shell distribution of HI and dust with constant density and temperature. Photons are injected centrally with the shell described by four physical parameters:

- $N_{HI}$ [ cm$^{-2}$ ] The radial HI column density
- $V_{exp}$ [ km s$^{-1}$ ] The expansion velocity of the shell
- $\tau_a$ [ dimensionless ] The radial optical depth due to dust absorption
- $b$ [ km s$^{-1}$ ] The Doppler parameter describing the microscopic HI velocity distribution

Model Lyα emission lines are generated a posteriori from arbitrary Lyα input spectra described by their full width at half maximum ($FWHM_{Ly\alpha}$) and $W_{Ly\alpha}$. $f_{esc}$ is computed for any model as the ratio of the transmitted Lyα flux to that of the input spectrum. With *MCLya* we have computed a large and complete grid of transfer models through shells covering the parameter space: log $N_{HI} \in [16—21.7]$ cm$^{-2}$; $V_{exp} \in [0—500]$ km s$^{-1}$; $\tau_a \in [0—5]$; $b \in [10—160]$ km s$^{-1}$. In total the grid consisted of 5,200 combinations of shell parameters with gridpoints approximately logarithmically spaced. Using the escape fractions of continuum photons near to Lyα and the same extinction law as applied to the observations[26], $\tau_a$ converts to the $E_{B-V} \in [0—0.36]$. To complete the range of models we generate spectra for a range of input $FWHM_{Ly\alpha} \in [50—700]$ km s$^{-1}$, giving us $f_{esc}$ and $E_{B-V}$ for approximately 50,000 synthetic galaxies.

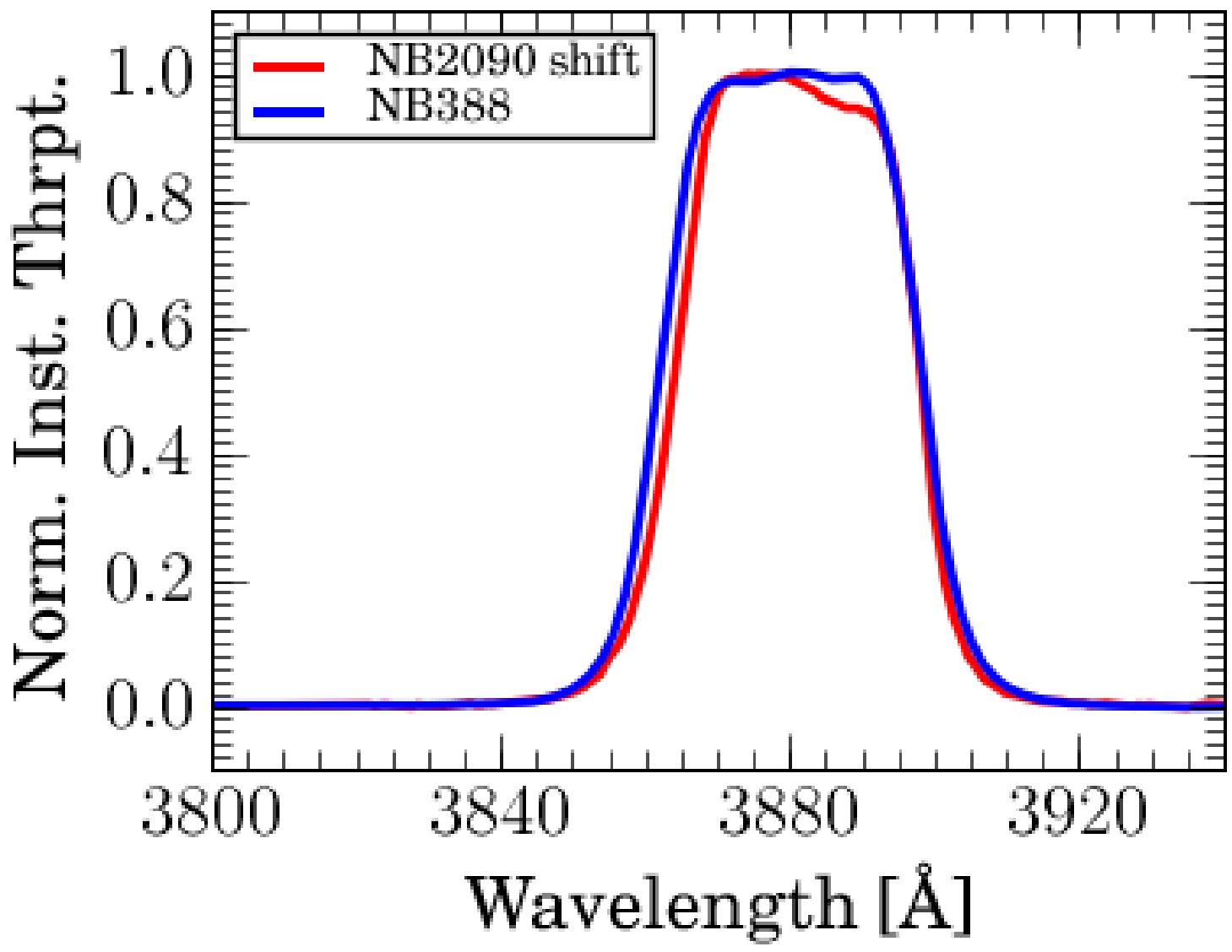

**Figure SI1:** The normalised instrumental response curves. The red line shows the *HAWK-I/NB2090* ("cosmological Hα") filter, re-scaled along the wavelength axis to sample Lyα (I.e. the theoretical perfect filter for Lyα). The blue curve shows the throughput of *FORS1/NB388*, which almost perfectly matches the defined set by the re-scaled *NB2090*.

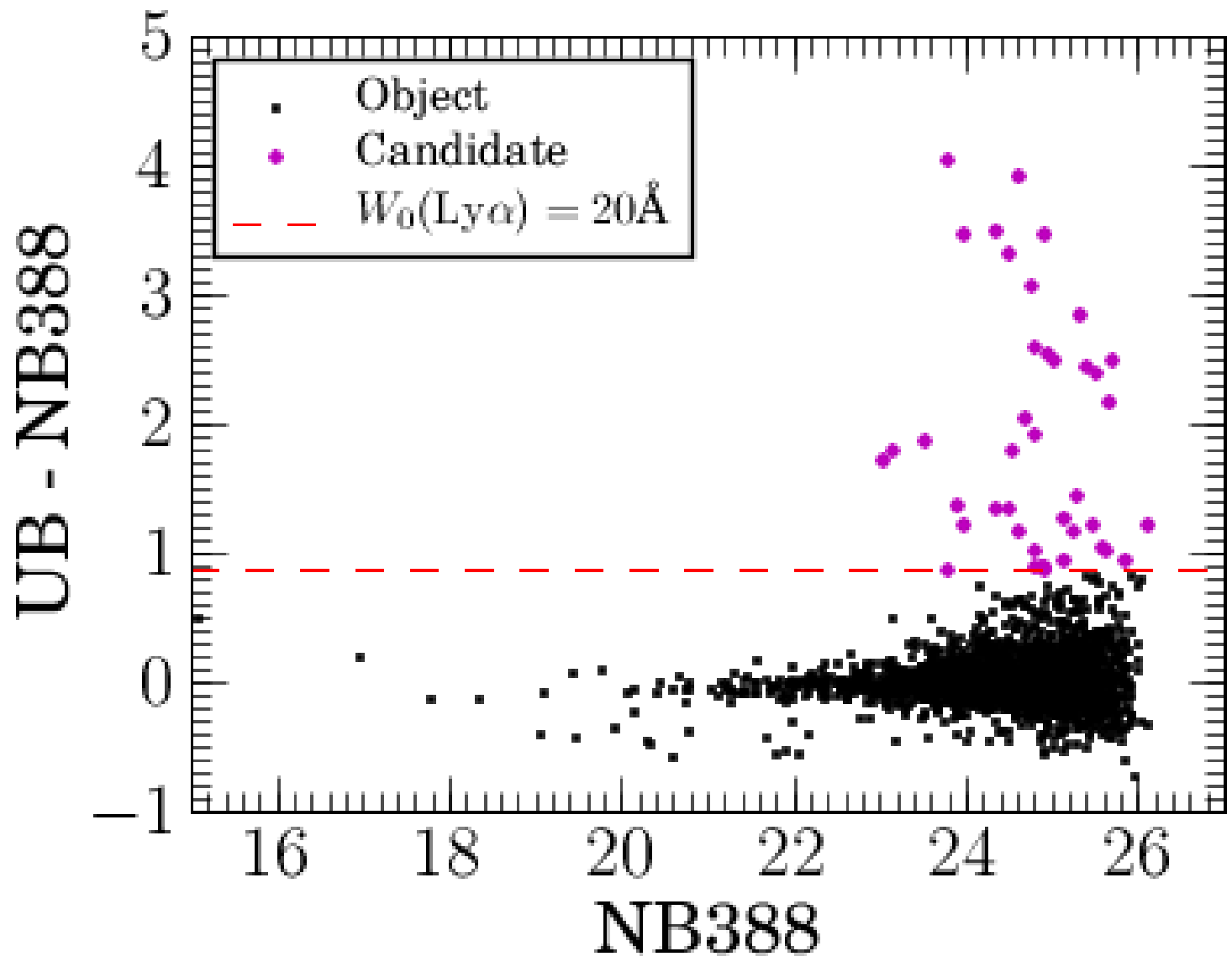

**Figure SI2.** The selection of narrowband excess candidates in colour—magnitude space. The dashed line corresponds to an equivalent width cut of 20Å in the $z$=2.2 restframe. Detections are represented by black spots, candidates by magenta.